\documentclass{cimento}

\usepackage{graphicx}  
\usepackage[utf8]{inputenc}   
  \usepackage[T1]{fontenc} 
\usepackage{amsmath}
\usepackage[style=numeric, sorting=none,citestyle=numeric-comp, doi=false,maxbibnames=1]{biblatex}
\renewbibmacro{in:}{}
\AtEveryBibitem{\clearfield{title}}

\newcommand{\im}{\operatorname{Im}}

\newcommand{\dzero}{D_0^*(2300)}
\newcommand{\dz}{D_0^*}
\newcommand{\done}{D_1(2430)}
\newcommand{\mev}{{\rm MeV}}

\title{Employing Approximate Symmetries for Hidden Pole Extraction}
\author{A.Asokan\from{ins:ju}\ETC\thanks{e-mail: a.asokan@fz-juelich.de}}
\instlist{\inst{ins:ju} Institute for Advanced Simulation and Institut für Kernphysik, Forschungszentrum Jülich, D-52425 J\"ulich, Germany,
}

\begin{document}

\maketitle
\begin{abstract}
Recent lattice analyses of the $D\pi$ scattering by Hadron Spectrum Collaboration(HadSpec) 
report only one pole in the $\dz$ channel. This is in odds with the unitarised chiral perturbation theory analyses, 
which predict the $\dzero$ as the interplay of two poles. We provide an explanation for this
contradiction~---~the exsistence of a hidden pole. We further show that the hidden pole 
can be better extracted from the lattice data by imposing SU(3) flavour constraints on the fitting 
amplitudes.
\end{abstract}

\section{Introduction}

From unitarized chiral perturbation theory analyses, the structure of $\dzero$ and $\done$ 
can be understood as the interplay of two poles, corresponding to two scalar/axial-vector isospin
doublet states with different SU(3) flavor content
~\cite{Kolomeitsev:2003ac,Guo:2006fu,Guo:2006rp,Guo:2009ct,Albaladejo:2016lbb,Du:2017zvv,Lutz:2022enz}. 
These states emerge from non-perturbative dynamics
of $D$ mesons scattering off the Goldstone boson octet. This two pole picture solves various problems 
that the experimental observation posed~\cite{Du:2017zvv}. However, in the recent lattice analysis from HadSpec of 
$D\pi-D\eta-D_s\bar{K}$ coupled channel system
at higher pion masses, only one pole was reported in the $\dz$ channel, while it was not possible to 
extract reliable parameters of a second pole from the lattice data~\cite{Moir:2016srx}. We here show that a higher pole does indeed exist
in the lattice data albeit on hidden sheets and propose the use of SU(3) flavour constrains to better capture its location

\section{Analytic Properties of the T-matrix}
\begin{figure}[h]
    \centering
    \includegraphics[height=0.40\textwidth]{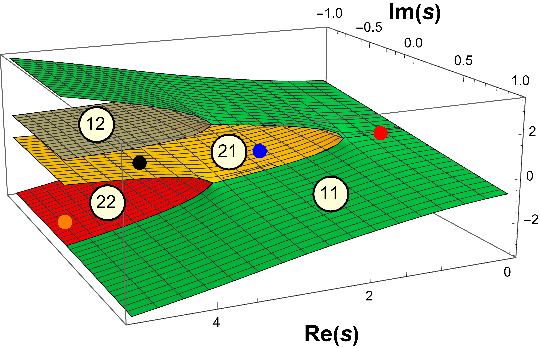}
    \caption{Illustration shows the sheet structure in case of a two channel case. The pole shown in 
    red correspond to a bound state while blue, black and orange represent resonances.}
    \label{fig:sheets}
\end{figure}

Except for branch cuts and poles the T-matrix is analytic over the whole complex plane. 
Right-hand cuts appear at every channel threshold, doubling the number of Riemann sheets.
The sign of the imaginary part of the centre of mass momentum($\im(p_{cm})$) is used to 
label the sheets; see tab.2 of ref.~\cite{Asokan:2022usm} for details. The physical sheet 
is the Riemann sheet in which $\im(p_{cm}) > 0$ for all channels. Depending on location, 
the poles correspond to bound states, 
virtual states or resonances. Hidden poles are resonance poles located on a sheet beyond the point at 
which it is connected directly to the physical sheet. Hence their effect on the amplitude can only be 
felt at the threshold, where the corresponding sheet is connected to the physical sheet.
For example, the blue and orange points in fig.~\ref{fig:sheets} are the poles which
would produce a 'conventional' resonance signal.
While the black point would correspond to a hidden pole. And its effect
on the amplitude would be felt at the second channel threshold. Poles located on the sheets closest to
the physical sheet have the largest influence on the amplitude.

\section{Analysis of the Lattice Study Amplitude}

The analysis in ref.~\cite{Moir:2016srx} was done by using a sizeable set of K-matrix 
parametrisations of the kind

\begin{align}
    K_{ij} = \frac{\left(g^{(0)}_i {+}\,g^{(1)}_i s\right)\left(g^{(0)}_j {+}\,g^{(1)}_j s\right)}{m^2 - s} 
    +\,\gamma^{(0)}_{ij} +\,\gamma^{(1)}_{ij} s,
    \label{eqn:Kdef}
\end{align}
where $i$ and $j$ label the various channels, $g^{(n)}_i$\,,$\gamma^{(n)}_{ij}$\ and $m$\ are real parameters determined 
from the fit to the lattice data. For the explicit form of the T-matrix see ref.~\cite{Moir:2016srx,Asokan:2022usm}. A pion mass of $m_{\pi} = 391 \mev$ was used 
in the lattice study.

\subsection{Pole Search}
In our labelling the closest sheets to the physical sheet
above the respective thresholds are the sheets labelled RS211, RS221 and RS222.
The lower pole in the studied channels appear as a bound state at the pion mass used in the lattice study
and is accordingly located on sheet RS111~\cite{Moir:2016srx}; the same was seen in 
UChPT~\cite{Albaladejo:2016lbb}.

In our analysis we find in addition to the reported bound state pole, additional higher poles in sheets RS211, 
RS221 and RS222. These additional poles were found in almost all parametrisations employed by HadSpec we analysed.
Figure~\ref{fig:plocfit} shows the locations of these higher poles in sheet RS221
from the HadSpec amplitude parametrisations and of the most prominent higher pole from the UChPT 
amplitude at the unphyical masses employed in the lattice study~\cite{Albaladejo:2016lbb}.

\begin{figure}[h]
    \centering
    \includegraphics[width=0.327\textwidth]{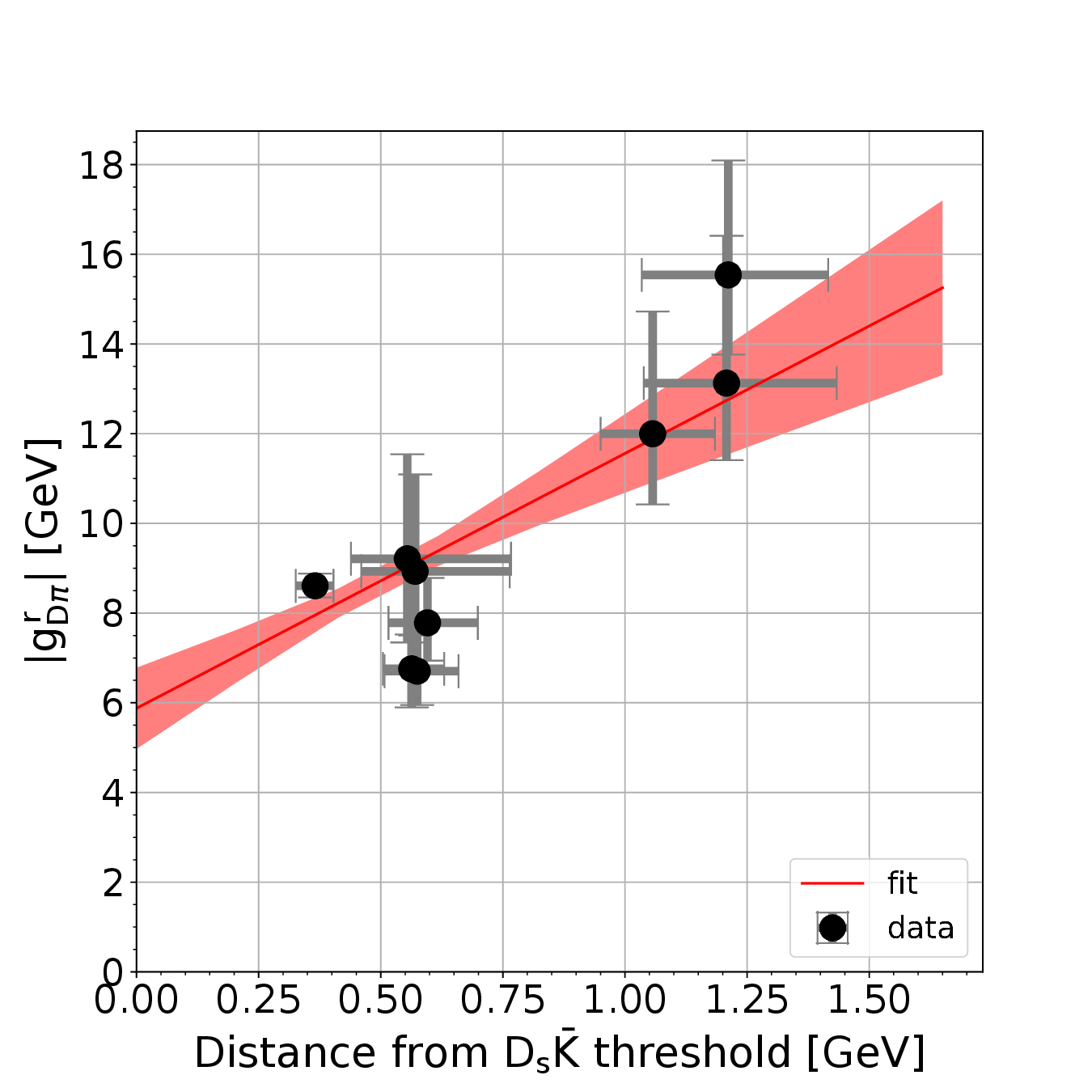}
    \includegraphics[width=0.327\textwidth]{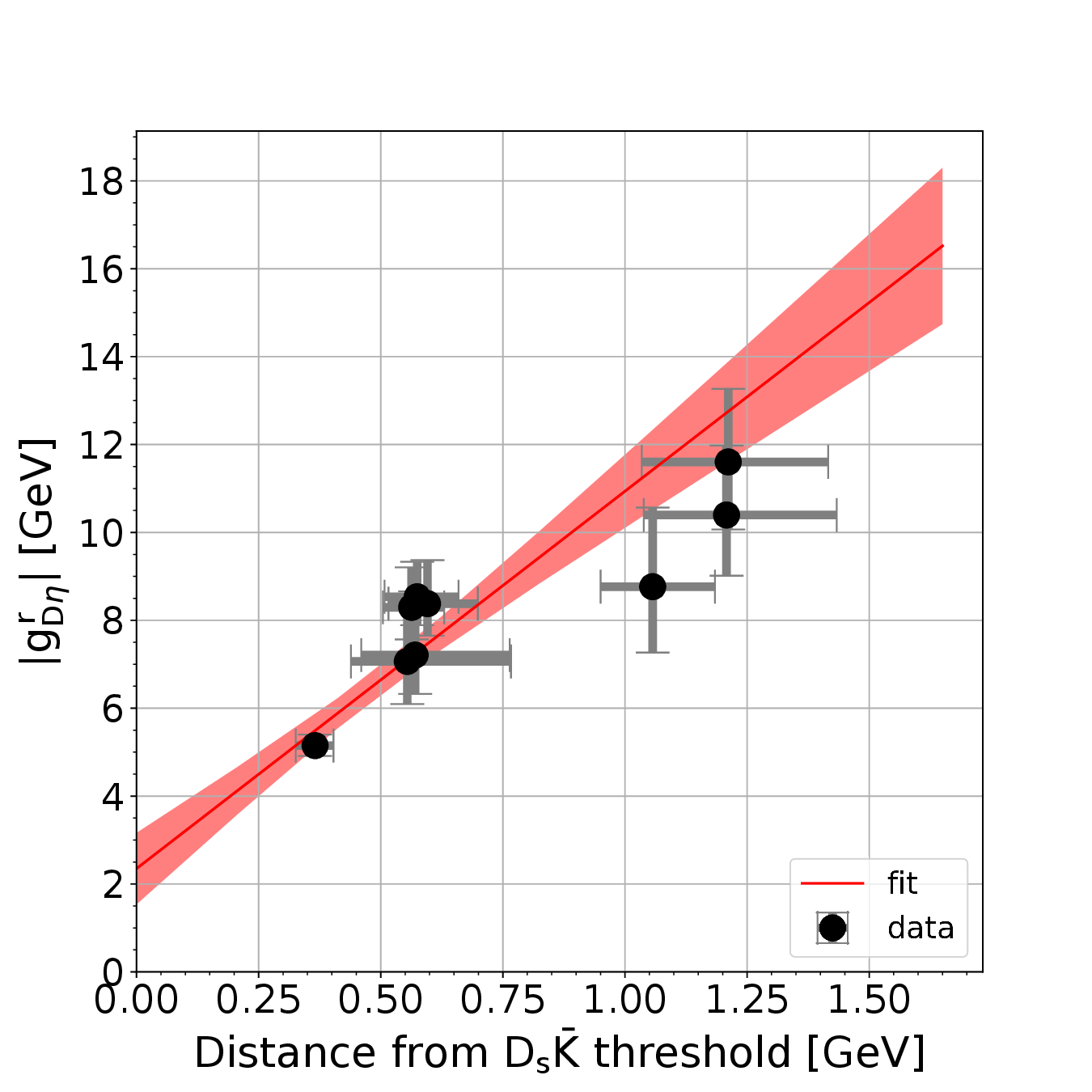}
    \includegraphics[width=0.327\textwidth]{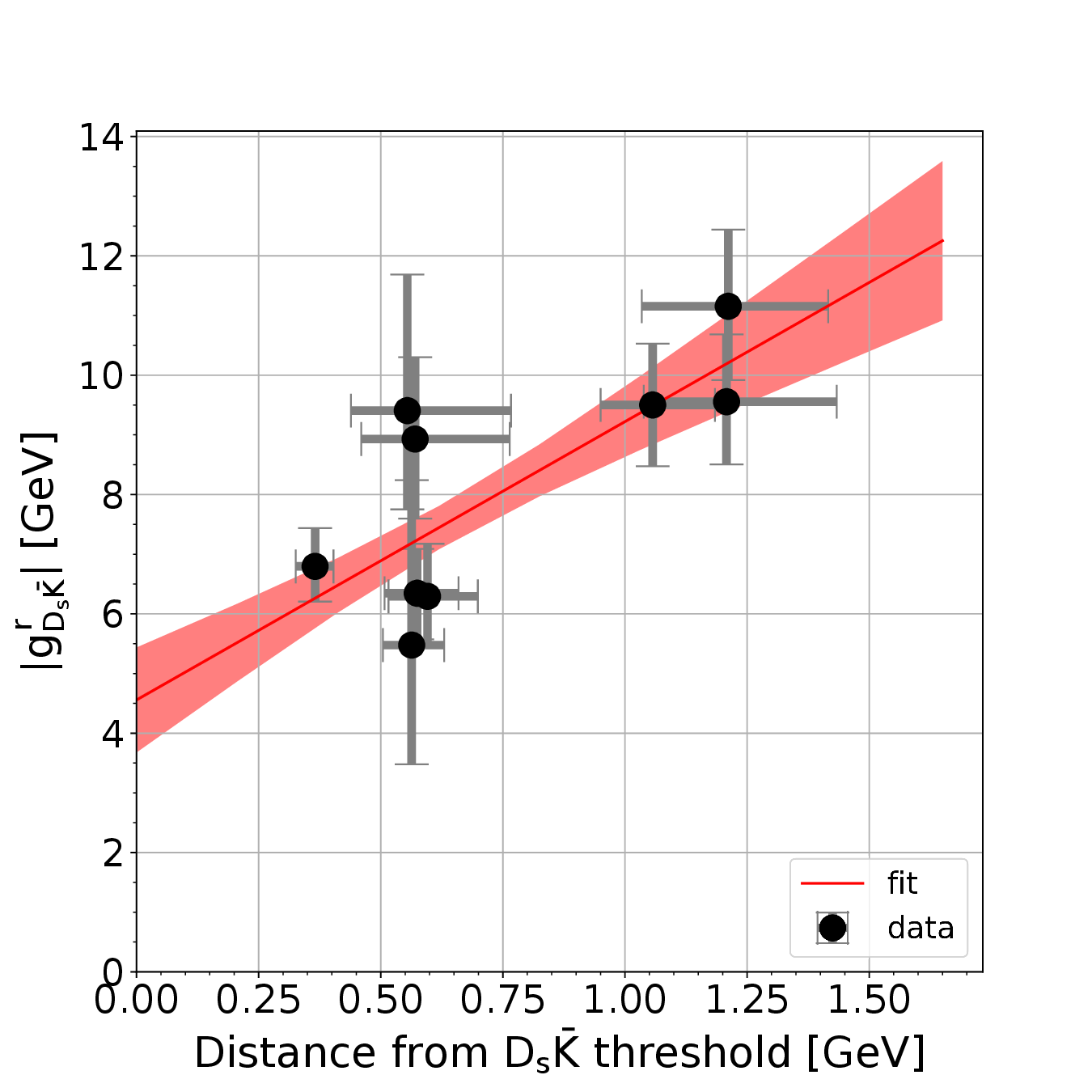}
    \caption{The distance of the real part of the pole from
    threshold versus the effective coupling of the pole to the 
    $D\pi$ channel (left), $D\eta$ channel (middle) and $D_s\Bar{K}$ channel (right). The red line shows the 
    straight line fit. The red band encloses the $1\sigma$\ uncertainty of the fit. 
    }
    \label{fig:resxdist}
\end{figure}

From fig.~\ref{fig:plocfit} we see that their locations do not scatter wildly between the 
parametrisations but show a significant correlation. All higher 
poles are seen to lie at higher energies than the point where their respective Riemann sheet connects to 
the physical sheet, \emph{i.e} they are located on hidden sheets. For example, the poles on RS221 lie 
beyond the $D_s\bar{K}$ threshold. They are shielded by RS222 and their effect can hardly be seen above 
the $D_s\bar{K}$ threshold. From the plot we see an increase in their widths as their distance from the 
threshold increases. We can see that the UChPT pole also lies in line with this correlation. To have an 
estimate of the effects of the poles at the respective threshold we plot the effective couplings from 
their residues to the different channels, with respect to the distance of the poles from the threshold. 
We quantified the distance as

\begin{align}
    {\rm Dist} = M - M_{\rm thr},
    \label{eqn:pdist}
\end{align}

where $M_{thr}$ is the relevant threshold and $M$ is the real part of the pole. Since the residues  
also drive the widths of the poles, eq.~\eqref{eqn:pdist} was chosen to quantify the distance.
Figure~\ref{fig:resxdist} shows the effective couplings \emph{vs} the distance plots. A straight line was 
fitted to extract the y~-~intercept to use as an estimate of the effect of the poles at threshold. From 
the plots it is seen that this value is well constrained. We interpret this as the 
lattice data requires, in addition to the bound state pole, a second pole. The ill constrained
second pole locations are being balanced by their enhanced residue. This mechanism was also observed
in the case of $f_0(980)$ and $a_0(980)$ ~\cite{Baru:2004xg}. This calls for a stronger constraint in the
amplitude parametrisation to better confine the location of the higher hidden pole.

\section{SU(3) Flavour Symmetry}

We put forward a SU(3) flavour constrained K-matrix to better capture the second pole location. 
Employing SU(3) flavour symmetry is justified as we work with higher pion masses which leads to
lower pion-kaon mass difference, moreover we are interested in the higher mass range of the second 
pole. In these circumstances the leading SU(3) breaking effects come from the loop functions.

The relation between the SU(3) flavour basis and the isopin-symmetric particle basis is given by
\begin{equation}
    \begin{pmatrix}
    \lvert [\Bar{3}] \rangle \\ 
    \lvert [6] \rangle \\
    \lvert [\overline{15}] \rangle 
    \end{pmatrix}
    = U
    \begin{pmatrix}
    \lvert D\pi \rangle \\ 
    \lvert D\eta \rangle \\
    \lvert D_{s}\Bar{K} \rangle 
    \end{pmatrix} ;
    \label{eqn:brel}
    U=
    \begin{pmatrix}
     -3/4 & -1/4 & -\sqrt{3/8}
   \\
   \sqrt{3/8} & -\sqrt{3/8} & -1/2
   \\
    1/4 & 3/4 & -\sqrt{3/8}
    \\
    \end{pmatrix}  .
\end{equation}

The flavour constrained K-matrix is of the form
\begin{align}
 K = \left(\frac{g_{\Bar{3}}^2}{m_{\Bar{3}}^2-s}{+} c_{\Bar{3}}\right)C_{\Bar{3}} + \left(\frac{g_{6}^2}{m_{6}^2-s}{+} c_{6}\right)C_{6} + 
 c_{\overline{15}}\,C_{\overline{15}}
 \label{eqn:su3k}
\end{align}
where matrices $C_{\Bar{3}}$, $C_{6}$ and $C_{\overline{15}}$ contain the SU(3) symmetric coupling 
strengths and can be read off from eq.~\eqref{eqn:brel}. For explicit forms see
eqs.(20, 21, 22) in ref.~\cite{Asokan:2022usm}. The free parameters are $g_\alpha$, $c_\alpha$ and
$m_\alpha$. Several fits with varying number of parameters were carried out to the energy levels in 
the lattice rest frame. The scheme introduced in ref.~\cite{Doring:2011vk} was followed to perform the
fits.

\begin{figure}[h]
    \centering
    \includegraphics[width=0.9\linewidth]{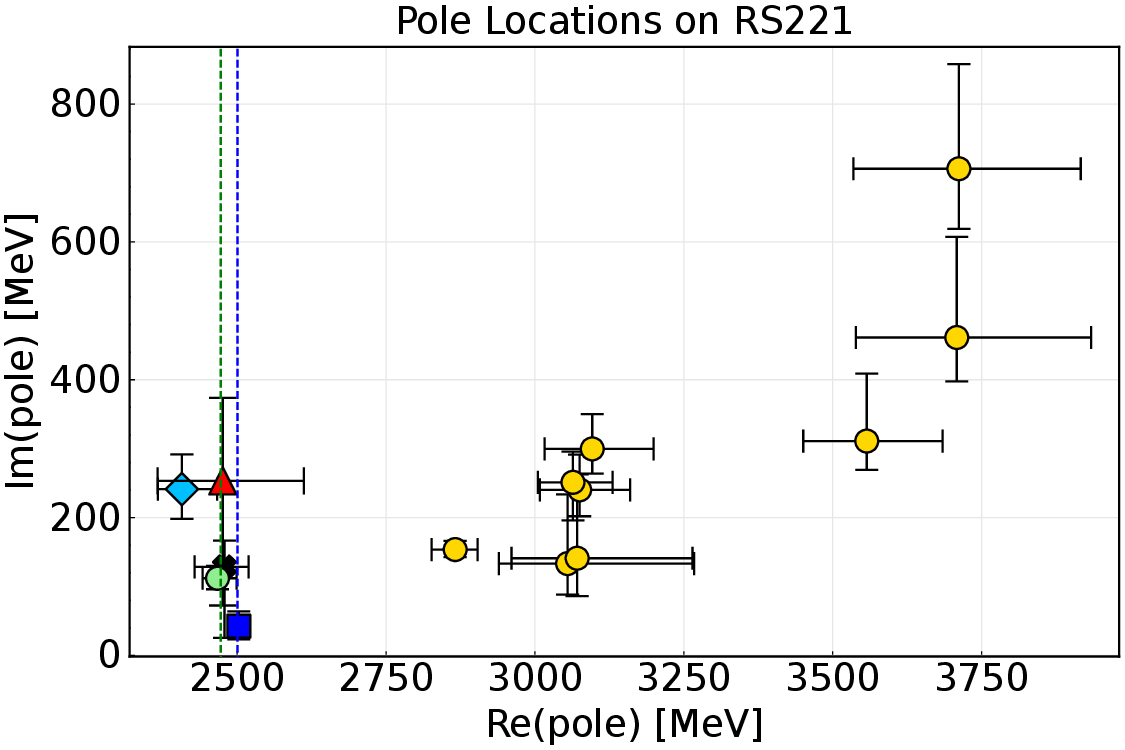}
    \caption{The location of poles on RS221 sheet. The $x$-axis and $y$-axis show the real and imaginary part of energy, respectively. The poles from the amplitude parametrizations employed in ref.~\cite{Moir:2016srx} are shown in yellow. The pole from the UChPT amplitude~\cite{Liu:2012zya} is shown in green~\cite{Albaladejo:2016lbb}.The higher pole locations extracted from the various fits are shown in red, cyan,black and blue. The vertical green and blue lines show the $D\eta$ and $D_s\bar{K}$ thresholds respectively.}
    \label{fig:plocfit}
\end{figure}

The higher pole locations extracted from fitting to the first four energy levels in the different
lattice volumes at rest are shown in fig.~\ref{fig:plocfit}. We see that
the flavour constrained amplitude is successful in better capturing the hidden higher pole.
The location of the lower bound state pole is consistent amongst all extractions. For a more 
in depth discussion on the quality of the fits and inclusion of higher lattice energy levels 
we refer the reader to ref.~\cite{Asokan:2022usm}.

\section{Conclusion}

A re-analysis of the K-matrix parametrisations provided in ref.~\cite{Moir:2016srx} found, in addition to 
the reported bound state pole, additional poles on unphysical Riemann sheets in every amplitude. Though 
their locations vary greatly between parametrisations, their effects on the amplitudes were comparable in 
all parametrisations. We identify the origin of this as their location on hidden sheets. In such scenarios 
their effect on the amplitude becomes visible at threshold, and their distance from threshold can be 
balanced by an enhanced residue. To extract the higher pole from lattice data we propose an SU(3) flavour 
constrain on the K-matrix. The flavour constrained amplitude is seen to be able to reproduce energy levels 
well and produce a pole at $D\eta$ and $D_s\bar{K}$ thresholds. It is interesting to obeserve that this 
pole is in fact consistent with that of the UChPT amplitudes. Such a symmetry constrained amplitude maybe 
used in analysing other lattice data and also experimental data involving multiple channels.

\acknowledgments
I am very grateful to Christopher Thomas and David Wilson for very valuable discussions and for providing us with the results of ref.~\cite{Moir:2016srx}. I thank all the co-authors of ref.~\cite{Asokan:2022usm}
for their valuable input and for their contributions to the analysis done therein.
This work is supported by the MKW NRW under the funding code NW21-024-A.

\printbibliography

@article{Kolomeitsev:2003ac,
    author = "Kolomeitsev, E. E. and Lutz, M. F. M.",
    title = "{On Heavy light meson resonances and chiral symmetry}",
    eprint = "hep-ph/0307133",
    archivePrefix = "arXiv",
    reportNumber = "GSI-PREPRINT-2003-20",
    doi = "10.1016/j.physletb.2003.10.118",
    journal = "Phys. Lett. B",
    volume = "582",
    pages = "39--48",
    year = "2004"
}

@article{Guo:2006fu,
    author = "Guo, Feng-Kun and Shen, Peng-Nian and Chiang, Huan-Ching and Ping, Rong-Gang and Zou, Bing-Song",
    title = "{Dynamically generated $0^+$ heavy mesons in a heavy chiral unitary approach}",
    eprint = "hep-ph/0603072",
    archivePrefix = "arXiv",
    doi = "10.1016/j.physletb.2006.08.064",
    journal = "Phys. Lett. B",
    volume = "641",
    pages = "278--285",
    year = "2006"
}

@article{Guo:2006rp,
    author = "Guo, Feng-Kun and Shen, Peng-Nian and Chiang, Huan-Ching",
    title = "{Dynamically generated $1^+$ heavy mesons}",
    eprint = "hep-ph/0610008",
    archivePrefix = "arXiv",
    doi = "10.1016/j.physletb.2007.01.050",
    journal = "Phys. Lett. B",
    volume = "647",
    pages = "133--139",
    year = "2007"
}

@article{Guo:2009ct,
    author = "Guo, Feng-Kun and Hanhart, Christoph and Mei{\ss}ner, Ulf-G.",
    title = "{Interactions between heavy mesons and Goldstone bosons from chiral dynamics}",
    eprint = "0901.1597",
    archivePrefix = "arXiv",
    primaryClass = "hep-ph",
    reportNumber = "FZJ-IKP-TH-2009-01, HISKP-TH-09-01",
    doi = "10.1140/epja/i2009-10762-1",
    journal = "Eur. Phys. J. A",
    volume = "40",
    pages = "171--179",
    year = "2009"
}

@article{Albaladejo:2016lbb,
    author = "Albaladejo, Miguel and Fernandez-Soler, Pedro and Guo, Feng-Kun and Nieves, Juan",
    title = "{Two-pole structure of the $D^\ast_0(2400)$}",
    eprint = "1610.06727",
    archivePrefix = "arXiv",
    primaryClass = "hep-ph",
    doi = "10.1016/j.physletb.2017.02.036",
    journal = "Phys. Lett. B",
    volume = "767",
    pages = "465--469",
    year = "2017"
}

@article{Du:2017zvv,
    author = "Du, Meng-Lin and Albaladejo, Miguel and Fern\'andez-Soler, Pedro and Guo, Feng-Kun and Hanhart, Christoph and Mei{\ss}ner, Ulf-G. and Nieves, Juan and Yao, De-Liang",
    title = "{Towards a new paradigm for heavy-light meson spectroscopy}",
    eprint = "1712.07957",
    archivePrefix = "arXiv",
    primaryClass = "hep-ph",
    doi = "10.1103/PhysRevD.98.094018",
    journal = "Phys. Rev. D",
    volume = "98",
    number = "9",
    pages = "094018",
    year = "2018"
}

@article{Lutz:2022enz,
    author = "Lutz, Matthias F. M. and Guo, Xiao-Yu and Heo, Yonggoo and Korpa, C. L.",
    title = "{Coupled-channel dynamics with chiral long-range forces in the open-charm sector of QCD}",
    eprint = "2209.10601",
    archivePrefix = "arXiv",
    primaryClass = "hep-ph",
    doi = "10.1103/PhysRevD.106.114038",
    journal = "Phys. Rev. D",
    volume = "106",
    number = "11",
    pages = "114038",
    year = "2022"
}

@article{Moir:2016srx,
    author = "Moir, Graham and Peardon, Michael and Ryan, Sin\'ead M. and Thomas, Christopher E. and Wilson, David J.",
    title = "{Coupled-Channel $D\pi$, $D\eta$ and $D_{s}\bar{K}$ Scattering from Lattice QCD}",
    eprint = "1607.07093",
    archivePrefix = "arXiv",
    primaryClass = "hep-lat",
    reportNumber = "DAMTP-2016-48",
    doi = "10.1007/JHEP10(2016)011",
    journal = "JHEP",
    volume = "10",
    pages = "011",
    year = "2016"
}

@article{Asokan:2022usm,
    author = "Asokan, Anuvind and Tang, Meng-Na and Guo, Feng-Kun and Hanhart, Christoph and Kamiya, Yuki and Mei\ss{}ner, Ulf-G.",
    title = "{Can the two-pole structure of the $D_{0}^{*}(2300)$ be understood from recent lattice data?}",
    eprint = "2212.07856",
    archivePrefix = "arXiv",
    primaryClass = "hep-ph",
    doi = "10.1140/epjc/s10052-023-11953-6",
    journal = "Eur. Phys. J. C",
    volume = "83",
    number = "9",
    pages = "850",
    year = "2023"
}

@article{Baru:2004xg,
    author = "Baru, V. and Haidenbauer, J. and Hanhart, C. and Kudryavtsev, Alexander Evgenyevich and Meissner, Ulf-G.",
    title = "{Flatte-like distributions and the $a_{0}(980)$ / $f_{0}(980)$ mesons}",
    eprint = "nucl-th/0410099",
    archivePrefix = "arXiv",
    reportNumber = "FZJ-IKP-TH-2004-19, HISKP-TH-04-22",
    doi = "10.1140/epja/i2004-10105-x",
    journal = "Eur. Phys. J. A",
    volume = "23",
    pages = "523--533",
    year = "2005"
}

@article{Doring:2011vk,
    author = "D{\"o}ring, M. and Meissner, Ulf-G. and Oset, E. and Rusetsky, A.",
    title = "{Unitarized Chiral Perturbation Theory in a finite volume: Scalar meson sector}",
    eprint = "1107.3988",
    archivePrefix = "arXiv",
    primaryClass = "hep-lat",
    doi = "10.1140/epja/i2011-11139-7",
    journal = "Eur. Phys. J. A",
    volume = "47",
    pages = "139",
    year = "2011"
}

@article{Liu:2012zya,
    author = "Liu, Liuming and Orginos, Kostas and Guo, Feng-Kun and Hanhart, Christoph and Meissner, Ulf-G.",
    title = "{Interactions of charmed mesons with light pseudoscalar mesons from lattice QCD and implications on the nature of the $D_{s0}^*(2317)$}",
    eprint = "1208.4535",
    archivePrefix = "arXiv",
    primaryClass = "hep-lat",
    reportNumber = "JLAB-THY-12-1599",
    doi = "10.1103/PhysRevD.87.014508",
    journal = "Phys. Rev. D",
    volume = "87",
    number = "1",
    pages = "014508",
    year = "2013"
}

\end{document}